# SOLAR DIFFERENTIAL ROTATION AND PROPERTIES OF MAGNETIC CLOUDS

K. Georgieva[1], B. Kirov[2], E. Gavruseva[3], J. Javaraiah[4,5]

1 - Solar-Terrestrial Influences Laboratory, Bulgarian Academy of Sciences, Bl.3 Acad.G.Bonchev str., 1113 Sofia, Bulgaria, E-mail: kgeorg@bas.bg
2 - Solar-Terrestrial Influences Laboratory, Bulgarian Academy of Sciences, Bl.3 Acad.G.Bonchev str., 1113 Sofia, Bulgaria, E-mail: bkirov@space.bas.bg
3 - Instituto Nazionale di Astrofisica, Osservatorio Astrofisico di Arcetri, Italy, E-mail: elena@arcetri.astro.it
4 - University of California, 430 Portola Plaza, Box 951547, Los Angeles CA 90095 USA, E-mail: jj@ucla.astro.edu
5 - Indian Institute of Astrophysics, Koramangala, Bangalore-560034, India, E-mail: jj@iiap.ernet.in

**ABSTRACT**

The most geoeffective solar drivers are magnetic clouds - a subclass of coronal mass ejections (CME's) distinguished by the smooth rotation of the magnetic field inside the structure. The portion of CME's that are magnetic clouds is maximum at sunspot minimum and mimimum at sunspot maximum. This portion is determined by the amount of helicity carried away by CME's which in turn depends on the amount of helicity transferred from the solar interior to the surface, and on the surface differential rotation. The latter can increase or reduce, or even reverse the twist of emerging magnetic flux tubes, thus increasing or reducing the helicity in the corona, or leading to the violation of the hemispheric helicity rule, respectively. We investigate the CME's associated with the major geomagnetic storms in the last solar cycle whose solar sources have been identified, and find that in 10 out of 12 cases of violation of the hemispheric helicity rule or of highly geoeffective CME's with no magnetic field rotation, they originate from regions with "anti-solar" type of surface differential rotation.

## 1. GEOEFFECTIVENESS OF CORONAL MASS EJECTIONS AND MAGNETIC CLOUDS

Richardson et al. (2001) studied the sources of geomagnetic storms over nearly three solar cycles (1972-2000) and found that the most intense storms as defined by Kp index at both sunspot minimum and sunspot maximum are almost all generated by coronal mass ejections (CME's). Our previous study (Georgieva and Kirov., 2005) demonstrated that the most geoeffective solar drivers are not CME's in general but magnetic clouds (MC's) – a subclass of CME's distinguished by enhanced magnetic field with smooth rotation inside the structure. Fig.1 and Fig.2 compare the geomagntic disturbances caused by an average CME and an average MC as expressed by the Kp and Dst indices, respectively. The figures are superposed epoch analyses of the daily average values of Kp and Dst indices on days of the event (day 0), one day before and after the event (days -1 and +1, respectively), etc. This study covers the period 1997-2002 in which we have 73 MC's defined by high magnetic field magnitude, low proton temperature or low plasma beta (ratio of the plasma pressure to the magnetic pressure), and smooth magnetic field rotation. We have used the list from Georgieva et al. (2005) completed by the events from http://cdaw.gsfc.nasa.gov/geomag_cdaw/. For the CME's we use the list of Richardson and Cane (2003) from which all events identified as MC's have been removed, which leaves us with a total of 128 cases.

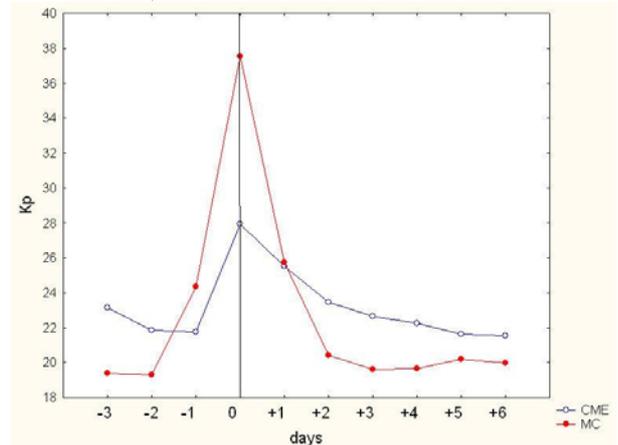

*Figure 1. Superposed epoch analysis of average daily Kp index on days with MC's and CME's in the period 1997-2002.*

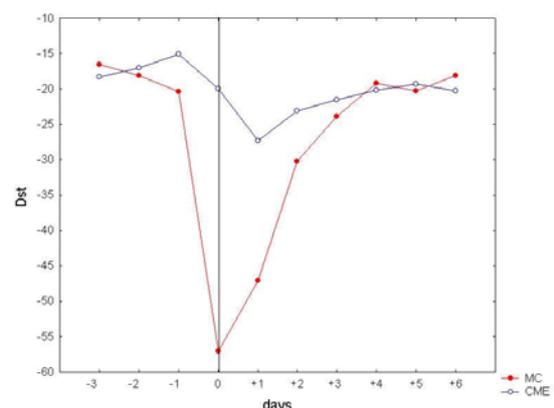

*Figure 2. Superposed epoch analysis of average daily Dst index on days with MC's and CME's in the period 1997-2002.*

## 2. HELICITY IN MAGNETIC CLOUDS

It has been noted that the occurrence frequency of CME's follows the sunspot cycle (Gopalswamy et al., 2003), while the occurrence frequency of MC's follows neither the sunspot cycle nor the occurrence frequency of CME's (Wu et al., 2003). Different estimates have been made about the percentage of CME's which are MC's: 30% (Gosling, 1990), 50% (Bothmer, 1996), 60-70% (Webb, 2000), until it was suggested that this ratio varies with the sunspot cycle – from practically 100% at sunspot minimum though with poor statistics, to 15% at sunspot maximum (Richardson and Cane, 2004). Our results (Georgieva and Kirov, 2005) confirm this conclusion (Fig.3). This explains why from sunspot min to sunspot max, the intensity of storms associated with CME's increases, however the degree of association between CMEs and storms decreases (Webb, 2000).

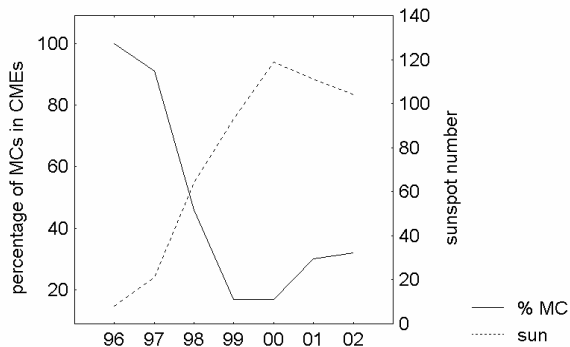

*Figure 3. Percentage of CME's which are MC's (solid line) and the sunspot cycle (dotted line); from Georgieva and Kirov, 2005.*

Magnetic helicity is constantly being generated in the Sun, but its dissipation is a very slow process (Berger, 1984), so the only way for the Sun to get rid of the helicity accumulated in the corona is to eject it in the CME's which carry this helicity away into the interplanetary space (Low, 1994). Therefore, the more helicity there is accumulated in the corona, the more of the CME's will contain helicity and will be registered as magnetic clouds. The amount of helicity in the corona depends on 2 factors: the net helicity transferred from the solar interior into the southern and northern corona (Berger and Ruzmaikin, 2000), and the surface differential rotation which can increase or reduce, or even reverse the twist of emerging magnetic tubes (De Vore, 2000).

### 2.1. Helicity transfer from the interior

The net helicity transferred from the solar interior into the northern corona is given by (Berger and Ruzmaikin, 2000):

$$\frac{dH_{CN}}{dt} = -\frac{\Omega_0}{2\pi}\Phi^2$$

where $H_{CN}$ is the net helicity from the solar interior into the northern solar hemisphere, $\Omega_0$ is the solar equatorial rotation rate (considered constant), and

$$\Phi = \pi R_V^2 B_1$$

is the net dipole flux through the northern photosphere. The expression for the southern hemisphere is with a positive sign. As the net dipole flux is maximum in sunspot minimum and minimum in sunspot maximum, the net transferred helicity has maximum positive values in the southern hemisphere and maximum negative values in the northern hemisphere around sunspot minimum, and is close to zero in both hemispheres around sunspot maximum (See Fig.3 in Berger and Ruzmaikin, 2000).

### 2.2. Surface differential rotation

In the corona, the emerging flux tubes are subjected to differential rotation (de Vore, 2000):

$$\Omega(\varphi) = \Omega_0 + b \sin^2\varphi$$

where $\Omega(\varphi)$ is the rotation rate at latitude $\varphi$, $\Omega_0$ is the equatorial rotation rate, and b is the latitudinal gradient of the rotation rate; **b**<0 so that the equator rotates faster than the higher latitudes. The generated helicity is negative in the north and positive in the south

$$H \sim -\text{sgn}(\varphi)\, \pi/32\, \cos(\varphi)$$

It twists initially untwisted or containing a finite amount of initial helicity flux tubes, and unwinds fields with opposite initial helicity.

## 3. VIOLATION OF THE HEMISPHERIC HELICITY RULE

The helicity is independent of the solar magnetic cycle and is always negative in the north and positive in the south (because $\Phi^2>0$). This hemispheric helicity rule was first noted by Seehafer (1990). The magnetic clouds carry the helicity of their source regions (Kumar and Rust, 1994). Therefore, MC's originating from the northern solar hemisphere, should have negative helicity (counterclockwise rotation of the magnetic field), and from the southern hemisphere – positive helicity (clockwise rotation). However, this is true in only 70-80% of the cases (Pevtsov et al., 1995).
We have studied 39 cases of major geomagnetic storms (Dst<-100) caused by CME's in the period 1997-2001 for which the solar sources of the CME's have been identified (http://cdaw.gsfc.nasa.gov/geomag_cdaw/). Out of them, in 27 cases (73%) a MC was observed at Earth's orbit with the expected hirality; In 7 cases

MC's originating from the northern solar hemisphere exhibited right-handed helicity (clockwise rotation of the magnetic field); In 3 cases MC's originating from the southern solar hemisphere exhibited left-handed helicity (counterclockwise rotation of the magnetic field), and in 2 cases CME's originating from the southern solar hemisphere were not MC's (no magnetic field rotation).

In 10 out of the 12 cases when the hemispheric helicity rule was violated, the solar differential rotation in the source region of the CME was "reversed" (increasing with latitude). The dates of the CME's, their identified source regions, the magnetic field rotation in the structure at the Earth's orbit, and the differential rotation in the source region are summarized in the Table.

*Table. CME's violating he hemispheric helicity rule*

| Date | Source region | CME at the Earth's orbit | rotation in source region |
|---|---|---|---|
| 23.11.97 | N20E05 | Right-handed MC | normal |
| 04.05.98 | S18E20 | No rotation | **reversed** |
| 14.11.98 | N18W02 | Right-handed MC | **reversed** |
| 12.02.00 | N25E26 | Right-handed MC | **reversed** |
| 14.10.00 | N01W14 | Right-handed MC | **reversed** |
| 18.04.00 | S20W58 | No rotation | **reversed** |
| 17.08.01 | N16W36 | Right-handed MC | normal |
| 26.09.01 | S12E23 | Left-handed MC | **reversed** |
| 01.10.01 | N10E18 | Right-handed MC | **reversed** |
| 03.10. 01 | S13E03 | Left-handed MC | **reversed** |
| 06.11.01 | N06W18 | Right-handed MC | **reversed** |
| 24.11.01 | S17W36 | Left-handed MC | **reversed** |

Fig.4 illustrates the normal solar rotation as derived from Solar Wilcox Observatory magnetic field measurements. The grid of the data used for the calculation of the rotation velocities is available online at http://wso.stanford.edu/synoptic. html and consists of 30 equal steps in sine latitude from 75.2 North to 75.2 South degrees and 5 degrees step in the heliographic longitude. The details of the calculations of the rotation velocity are described by Gavryuseva and Godoli (2005).

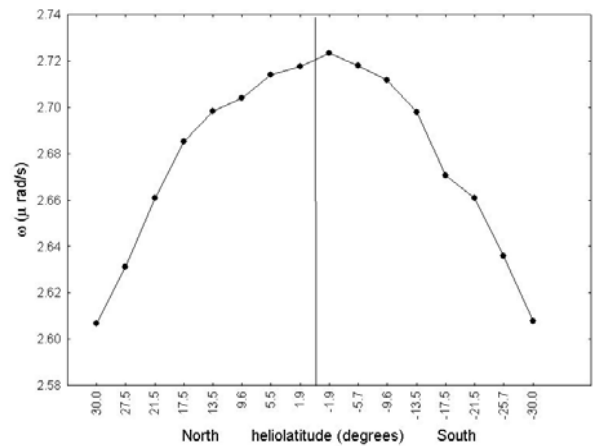

*Figure 4. Dependence of the rotation rate on latitude averaged over 333 Carrington rotations from 1976 to 2001 (from magnetic field measurements)*

Fig.5 demonstrates a case of differential rotation reversed in the source region of a CME (marked by a circle).

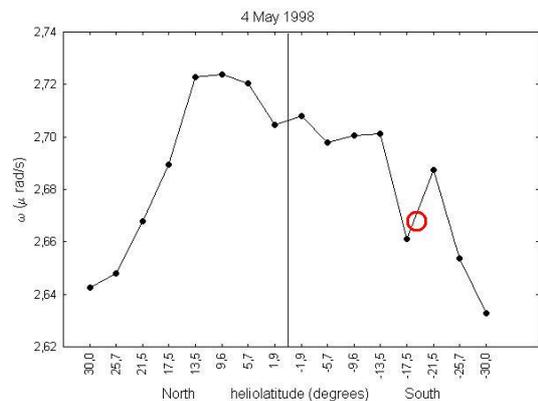

*Figure 5. An example of reversed differential rotation in the source region of a CME*

### 4. REVERSED (ANTI-SOLAR) DIFFERENTIAL ROTATION

The origin and maintenance of the solar differential rotation is usually explained by the interaction between rotation and convection (Kitchatinov, 2005). Convective turbulence in rotating medium is subjected to the Coriolis force, and the feedback distorts rotation and makes it nonuniform. Models for differential rotation based on this theory provide rotation law for the Sun (Kitchatinov & Rűdiger, 1995) in close agreement with helioseismology (Schou et al., 1998). On the other hand, the theory of the reversed, or "anti-solar" differential rotation, is still in its initial phase. The suggestion is that this type of rotation can result from a fast meridional flow which in turn can be caused by deviation from the spherical symmetry in the gravity or temperature distributions. The reason can be large-scale thermal inhomogeneities or tidal forcing from a companion star. About ten stars have already been identified with anti-solar type of differential rotation (Strassmeier et al., 2003), six of them are close binaries, and one is a giant for which dark spots have been observed at low latitudes with temperature contrast of about 200° K (Kitchatinov & Rűdiger, 2004).

In the case of the Sun, the anti-solar type of rotation is only observed in narrow latitudinal zones and in limited temporal intervals. In some of the cases listed in the table above, the planetary configurations hint at a possibility for tidal forcing. However, much additional study is needed to confirm this suggestion, completed by detailed data for the meridional circulation and the temperature distribution on the Sun.

## 5. SUMMARY AND CONCLUSIONS

The portion of CME's which are magnetic clouds is determined by the amount of helicity transferred from the solar interior into the corona, and on the surface differential rotation. The helicity transferred from the solar interior into the corona is always positive in the southern solar hemisphere and negative in the northern hemisphere, irrespective of the magnetic polarity cycle. The surface differential rotation can additionally wind or unwind the rising magnetic flux tubes, depending on the type of rotation – "solar type" or "anti-solar type". The cases of "anti-solar type" of differential rotation are related to violation of the hemispheric helicity rule. "Anti-solar type" of differential rotation can be caused by planetary alignments.